\documentclass[twocolumn,nofootinbib,showkeys,reprint]{revtex4-1}
\usepackage{amsmath,amssymb,amsthm,bm,braket,ascmac,url,hyperref,color,comment} 
\usepackage[pdftex]{graphicx}
\begin{document}
\title{An Approximation Scheme for Multivariate Information\\ based on Partial Information Decomposition}
\author{Masahiro Takimoto}
\email{masahiro.takimoto0618@gmail.com}
\affiliation{\vspace{2mm}Linea Co.,Ltd., Japan}

\date{\today}
\begin{abstract}
We consider an approximation scheme for multivariate
information assuming that synergistic information only
appearing in higher order joint distributions is suppressed,
which may hold in large classes of systems.
Our approximation scheme gives a practical way
to evaluate information among random variables
and is expected to be applied to
  feature selection in machine learning.
The truncation order of our approximation scheme
is given by the order of synergy. 
In the classification of information,
we use
the partial information decomposition
of the original one.
The resulting multivariate information is 
expected to be reasonable if higher order synergy is
suppressed in the system.
In addition, it is calculable 
in relatively easy way if the truncation order is not
so large. 
We also perform numerical experiments to check
the validity of our approximation scheme.

\end{abstract}
\maketitle

\section{Introduction}\label{sec:intro}
Mutual information~\cite{doi:10.1002/j.1538-7305.1948.tb01338.x}
is one of the fundamental measures which
captures information between two sectors. 
Even when each sector
contains a lot of random variables, the mutual information
can give total amount of information between two sectors.
One natural question would be how such information is distributed
among variables.
The framework called the partial information decomposition~\cite{DBLP:journals/corr/abs-1004-2515}
provides a way to decompose mutual information into
combinations of partial information in the form of unique,
redundant and synergistic information.

In practical point of view, the mutual information may have
some difficulties in multivariate and small sample cases.
For example, when the number of samples is much smaller than
that of possible realizations, it is hard to estimate the mutual information precisely. This is because the mutual information
depends on joint probability with a lot of arguments and
such a probability is hard to estimate in small sample cases.
Thus, it might be good to have an approximation scheme for 
mutual information that overcome some difficulties.

When we construct an approximation scheme, we need to
identify a kind of small quantities in the system.
In the multivariate information, one natural assumption
would be that information only appearing in higher order
synergies is suppressed. Though this assumption would be expected to hold in large classes of systems,
we have to specify what is 
information in higher order synergy.
The framework of partial information decomposition
gives one solution of this specification.

In this paper, we consider an approximation scheme for
mutual information with a lot of variables.
We rely on the assumption of suppression of information in higher order parts and construct an approximation scheme based on
the partial information decomposition.
The resulting approximation scheme is expected to be
reasonable if information in higher order synergies 
is suppressed. 
In addition, it is calculable in practice when truncation order is 
not so large.

This paper is organized as follows.
In sec.~\ref{sec:opid}, we show an overview of the partial information decomposition. In sec.~\ref{sec:ass}, we derive our approximation
scheme for mutual information. In sec.~\ref{sec:num},
we perform some numerical experiments to see the validity
of our scheme.
Sec.~\ref{sec:dis} is devoted to discussions.

\section{Overview of Partial Information Decomposition}\label{sec:opid}
Let $X_1,..,X_N,Y$ be random variables. For simplicity, we assume all random variables take discrete values and have finite state spaces.
Throughout this paper, we regard $X_1,..,X_N$ as feature variables and $Y$ as a target variable.
For a given set of feature variables $\{X_{i_1},..,X_{i_m}\}$, the mutual information on the target
variable $Y$ is defined to be
\footnote{In this paper, we denote mutual information with
a lot of feature variables as multivariate information or simply, mutual information.}
\begin{align}
MI&(X_{i_1},..,X_{i_m}:Y)\nonumber \\
&=\sum_{x_{i_1},..,x_{i_m},y}p(x_{i_1},..,x_{i_m},y)
\log\frac{p(x_{i_1},..,x_{i_m},y)}{p(x_{i_1},..,x_{i_m})p(y)},
\end{align}
where $p$ denotes probability and small letters indicate values of corresponding 
random variables.
The mutual information measures the amount of information about the target
variable $Y$ contained in selected features.
Though the mutual information can give us the total amount of information in features,
it might not be clear how the information is distributed among features $X_{i_1},..,X_{i_m}$.

\begin{figure}[h!]
\centering
\includegraphics[width=0.74\hsize]{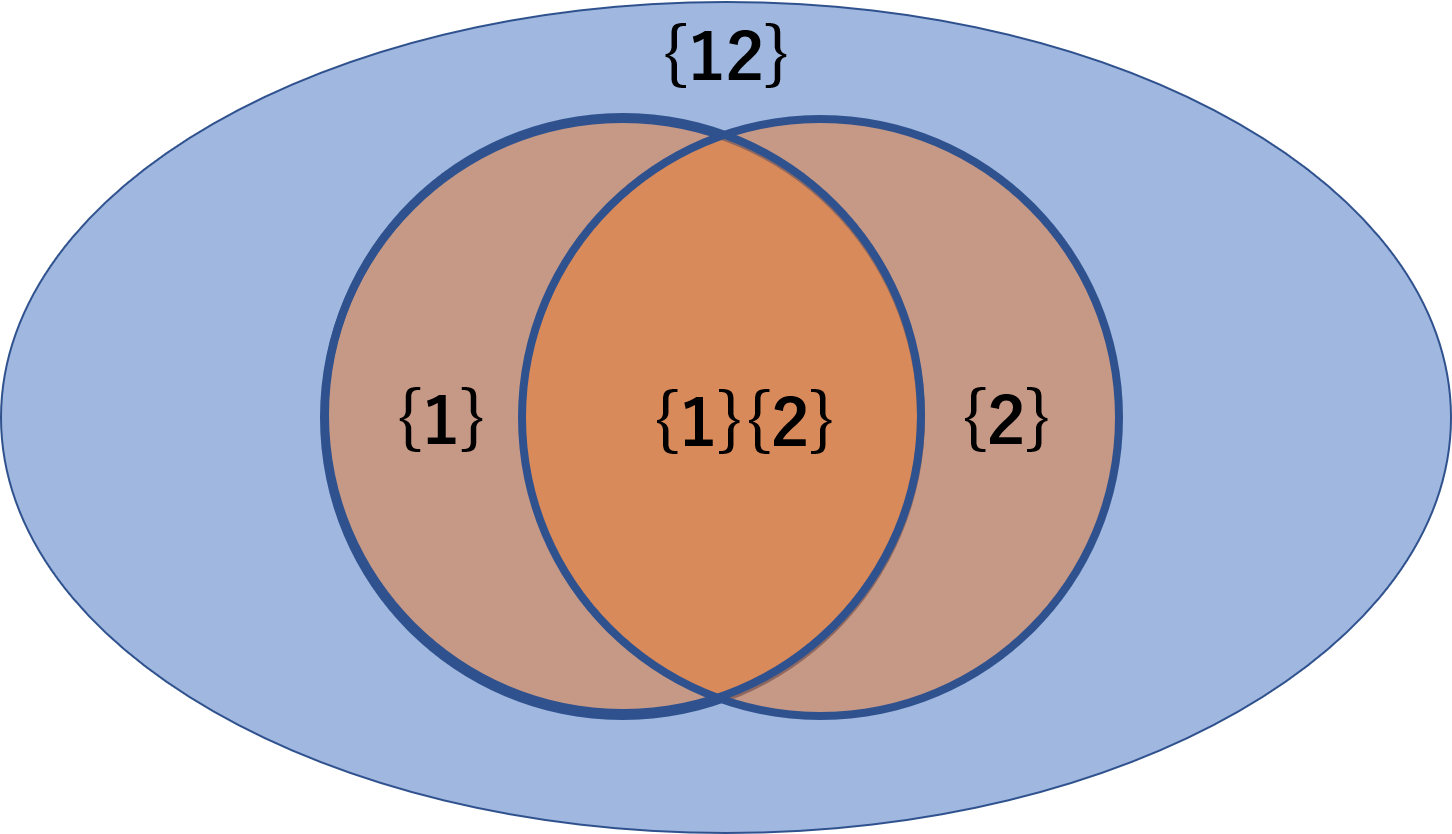}
\caption{
A partial information diagram for two feature variable case.
}
\label{fig:two}
\end{figure}

\begin{figure}[h!]
\centering
\includegraphics[width=0.74\hsize]{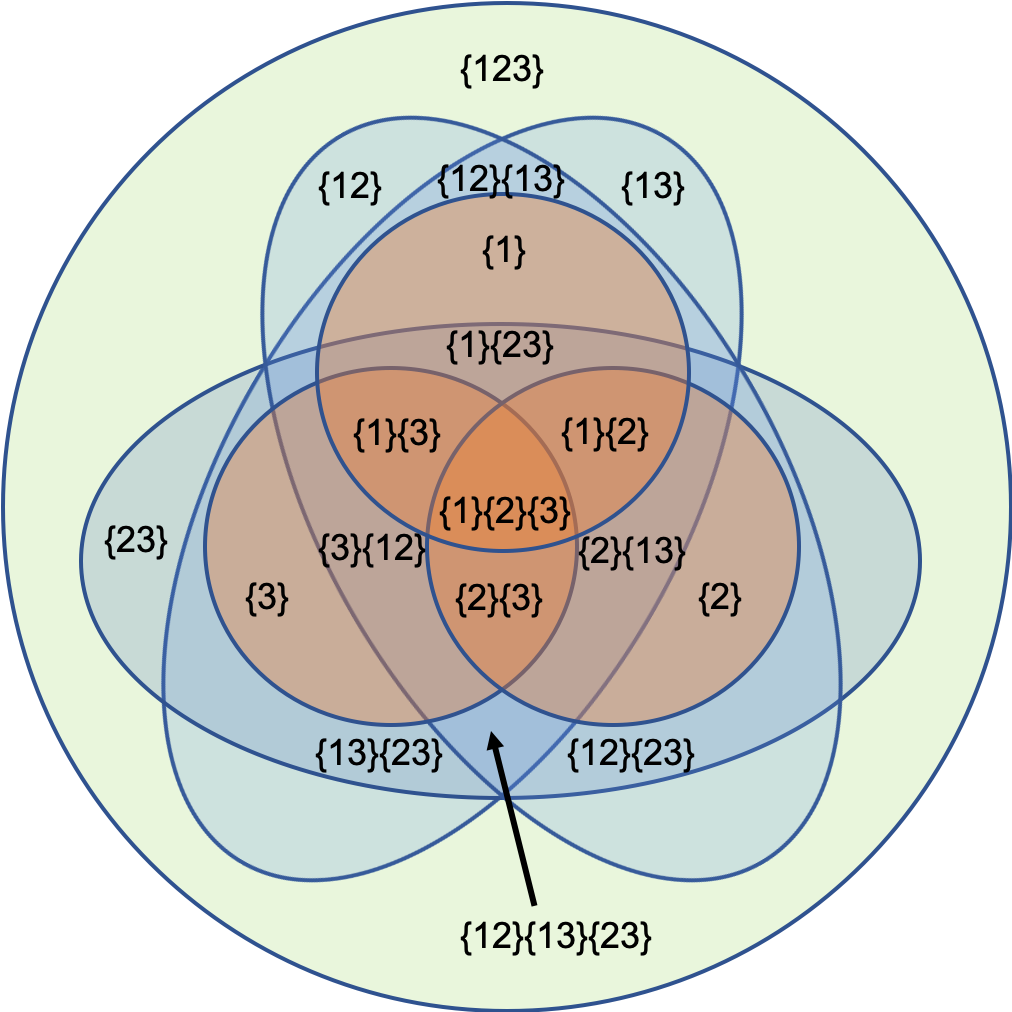}
\caption{
A partial information diagram for three feature variable case.
}
\label{fig:three}
\end{figure}

The framework of partial information decomposition (PID) can decompose the total amount
of information and gives how the information is distributed.
Here, we briefly show an overview of this framework and refer
to~\cite{DBLP:journals/corr/abs-1004-2515} for more details.
In PID framework, the total amount of information is decomposed into
unique, redundant and synergistic information that are associated with combinations of feature variables.
Fig.~\ref{fig:two} shows decomposition for two feature variable case $X_1,X_2$ and
this kind of diagram is denoted as partial information diagram.
The region $\{12\}$ is corresponding to synergistic information between $X_1$ and $X_2$.
The region $\{1\}$ or $\{2\}$ is corresponding to
unique information of $X_1$ or $X_2$ respectively.
The region $\{1\}\{2\}$ is redundant information
between $X_1$ and $X_2$.
Similarly, fig.~\ref{fig:three} indicates the information decomposition of three feature variable case $X_1,X_2,X_3$.
In this figure,
the mutual information is also decomposed into combinations of feature variables. For example, the region $\{3\}\{12\}$ is
corresponding to the unique part of the redundant information between $X_3$ and joint variable $(X_1,X_2)$.

The key function to determine this kind of information decomposition is the redundancy
information $I_\cap(Y:A_1,..,A_k)$ that measures information about $Y$ contained
in all $A_1,..,A_k$, where each $A_i$ is a subset of $\{X_1,..,X_N\}$.
Once $I_\cap$ is defined, every piece of partial information is determined.
In the original paper~\cite{DBLP:journals/corr/abs-1004-2515}, the redundancy information is denoted as $I_{\rm min}$
and defined to be
\begin{align}
\label{eq:cap}
I_{\rm min} (Y:A_1,..,A_k)=\sum_yp(y)\min_{A_i}I(Y=y:A_i),
\end{align}
where $I(Y=y,A_i)$ measures the information associated with a 
given outcome $y$ of $Y$:
\begin{align}
I(Y=y:A_i)=\sum_{a_i}p(a_i|y)\left[
\log \frac{1}{p(y)}-\log \frac{1}{p(y|a_i)}
\right].
\end{align}
This redundancy information $I_{\min}$ satisfies some basic axioms and has 
good features. For example, each part of partial information is
ensured to be non-negative.
However, it is known that $I_{\min}$ sometimes gives unintuitive results and several alternatives have been discussed and proposed in the literature in order to overcome drawbacks~\cite{DBLP:journals/corr/abs-1303-3440,DBLP:journals/corr/abs-1210-5902,DBLP:journals/corr/abs-1207-2080,DBLP:journals/corr/abs-1205-4265,DBLP:journals/corr/BertschingerROJA13,DBLP:journals/corr/RauhBOJ14,jisaku1,DBLP:journals/corr/PerroneA15,DBLP:journals/corr/RosasNEPV15,DBLP:journals/corr/GriffithCJEC13,DBLP:journals/corr/GriffithH14,DBLP:journals/corr/QuaxHS16,DBLP:journals/corr/Barrett14,Chicharro2017QuantifyingMR,Finn2018PointwisePI,Kolchinsky2019ANA,Ay2019InformationDB,Sigtermans2020API}.
Nevertheless,
in practical point of view, $I_{\rm min}$ would be still attractive because
the structure is relatively simple and it requires relatively small computational costs.
In this paper, we use $I_{\rm min}$ to define an approximation scheme for
multivariate information.

For later convenience, we consider union information $I_\cup(Y:A_1,..,A_k)$,
that measures information contained in any of $A_i$.
For simplicity, we use a shorthand notation $I_\cup(Y,\{A_i\})\equiv I_\cup(Y:A_1,..,A_k)$.
The principle of inclusion and exclusion gives

\begin{align}
I_\cup (Y:\{A_i\})&=\sum_{i}I_{\rm min}(Y:A_i)
-\sum_{i<j}I_{\rm min} (Y:A_i,A_j) \nonumber \\
&+\sum_{i<j<k}I_{\rm min} (Y:A_i, A_j , A_k)\cdots.
\end{align}
For further computations, the maximum-minimum identity is useful.
The identity states that for a given set of numbers $B=\{b_1,b_2,..\}$, we have
\begin{align}
{\rm max}~B&=\sum_i {\rm min}(b_i)-\sum_{i<j}{\rm min}(b_i,b_j)\nonumber \\
&+\sum_{i<j<k}{\rm min}(b_i,b_j,b_k)+\cdots.
\end{align}
With this identity, we obtain a simple expression of $I_\cup$:
\begin{align}
\label{eq:cup}
I_\cup (Y:\{A_i\})&=\sum_{y}p(y)\max_{A_i}I(Y=y:A_i).
\end{align}

\section{An Approximation scheme in terms of synergistic order}\label{sec:ass}

In practical point of view, a mutual information with a lot of feature variables
would have some disadvantages. For example, it would require a lot of computational costs.
In addition, if sample size is not so large, 
it would be hard to estimate
mutual information precisely.
Thus, it might be good to have a reasonable approximation which overcomes some of these
disadvantages.

In order to construct an approximation scheme, we have to specify 
a kind of small quantities in the system based on some reasonable assumption.
In our case,
one natural assumption might be that information only appearing in higher order
synergistic part is assumed to be small.
This assumption leads us to an approximation scheme whose accuracy 
is verified by smallness of higher order synergies.
Here, we denote the order of synergy as a number of joint features involved.
For example, the order of synergy for the element $\{23\}$ in fig.~\ref{fig:three} is two.

\begin{figure}[h!]
\centering
\includegraphics[width=0.74\hsize]{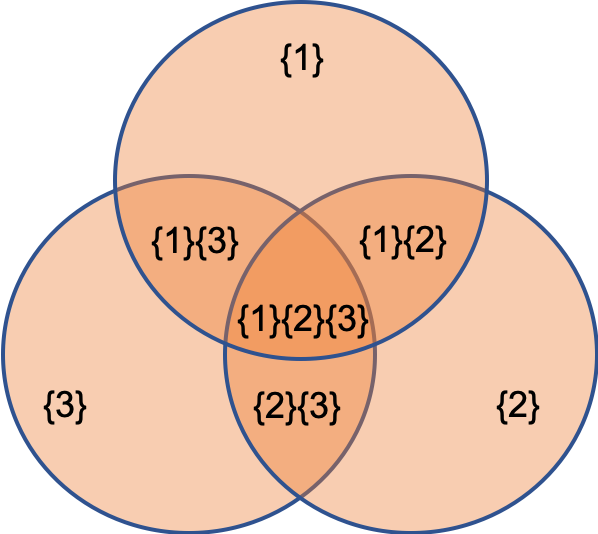}
\caption{
Information contained in $I^{(1)}$ for three feature variable case.
}
\label{fig:i1}
\end{figure}

\begin{figure}[h!]
\centering
\includegraphics[width=0.74\hsize]{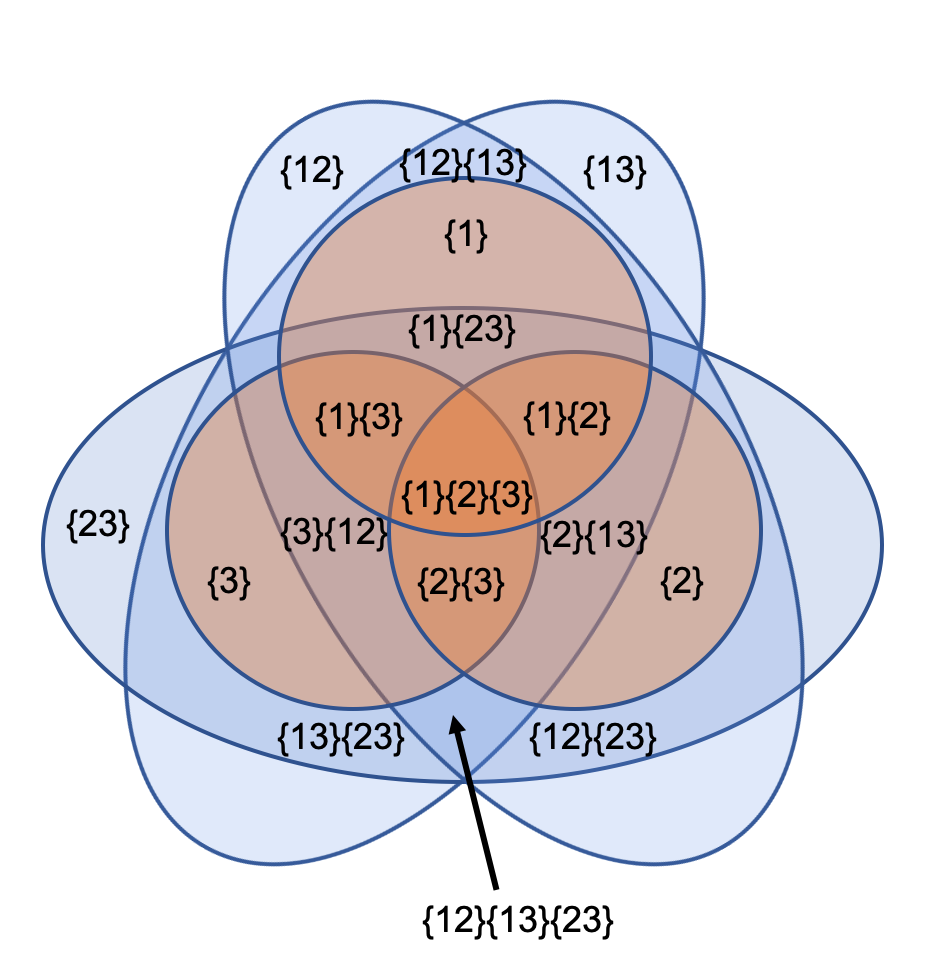}
\caption{
Information contained in $I^{(2)}$ for three feature variable case.
}
\label{fig:i2}
\end{figure}

Next, let us relate the order of synergy to an approximation scheme.
We denote a set of set of feature variables that contains just $n$ types of features as $C^{(n)}$.
For example, we have $C^{(1)}=\{\{X_1\},\{X_2\},...\}$ and $C^{(2)}=\{\{X_1,X_2\},\{X_1,X_3\},..\}$.
We define $I^{(k)}\equiv I_\cup (Y:C^{(k)})$ as the total amount of information
that takes synergistic information 
up to $k$ joint features 
into account.
For example,
$I_\cup(Y:C^{(1)})$ could be interpreted as the total amount of information
without any synergy between feature variables.
$I_\cup(Y:C^{(2)})$ could be regarded as the total amount of information
that takes any pairs of synergistic information into account.
$I^{(N)}$ is corresponding to the multivariate information
of whole feature variables $X_1,..,X_N$.
The difference
\begin{align}
\Delta^{(k+1)}=I^{(N)}-I^{(k)},
\end{align}
would be interpreted as the total amount of information that only appears
in the synergistic information involved by more than $k$ features.
Fig.~\ref{fig:i1} and fig.~\ref{fig:i2} are corresponding to
information contained in $I^{(1)}$ and $I^{(2)}$ respectively
for three variable case. 
By using Eq.~(\ref{eq:cup}), the quantity $I^{(k)}$ can be written as
\begin{align}
\label{eq:ik}
I^{(k)}=\sum_{y}p(y)\max_{C_i^{(k)} \in C^{(k)}}
I(Y=y:C_i^{(k)}),
\end{align}
where $C_i^{(k)}$ denotes an element in the set $C^{(k)}$.

The quantity $I^{(k)}$ has following features:
\begin{itemize}
\item Each $I^{(k)}$ has a corresponding region in the partial information diagram.
\item $I^{(k)}$ is increasing function in terms of $k$
\item $I^{(k)}$ only depends on joint probability functions whose number of
arguments are $k+1$: ($p(y,x_{i_1},..,x_{i_k})$).
\item Lower order $I^{(k)}$ is expected to be stable against small sample size
because it does not depend on joint probabilities with large number of arguments.
\item If higher order synergistic information is small, low order $I^{(k)}$ is
expected to give a reasonable approximation of the total mutual information.
\end{itemize}
Considering these features,
collection of 
$I^{(k)}$ might be regarded as an approximation scheme for the total mutual information.

Finally, let us comment on the feature selection using $I^{(k)}$.
The feature selection based on mutual information
seems promising and
 a lot of methods has been derived in the literature (see for
 example~\cite{DBLP:journals/corr/VergaraE15,jisaku3,DBLP:journals/corr/abs-1907-07384}).
The feature selection based on $I^{(k)}$ would become a new one. 
One simple way to determine important features based on
$I^{(k)}$ may
be as follows.
For a given order $k$ that is not so large,
we can estimate $I^{(k)}$ for all features.
Then,
Some of features may not contribute to $I^{(k)}$.
This is 
because 
number of 
features that is chosen in max operation in Eq.~(\ref{eq:ik})
is limited.
We delete such irrelevant features and obtain
a set of features that is relevant in $I^{(k)}$.
If number of features is still large,
we may delete additional features 
 by referring to $I^{(k)}$ with lower number of features.

\section{Numerical Results}
\label{sec:num}
In this section, we perform two numerical experiments in a simple setup. The first experiment is intended to see
behavior of exact $I^{(k)}$. The second one is
to see effects of finite sample size.

Let $s_1,s_2..,s_M$ be bit type random variables whose values are $0$ or $1$.
We set the joint probability $p(s_1,..,s_M)$ as follows.
\begin{align}
p(s_1,..,s_M)&=\frac{e^{A(s_1,..,s_M)}}{Z},\\
Z&=\sum_{s_1,..,s_M}e^{A(s_1,..,s_M)},\\
A(s_1,..,s_M)&=
\epsilon_0\sum_i a_i\cdot s_i+\epsilon_1\sum_{i<j}b_{ij}\cdot(s_i\oplus s_j)\nonumber \\
&+\epsilon_2\sum_{i<j<k}c_{ijk}\cdot(s_i\oplus s_j\oplus s_k),
\end{align}
where $\oplus$ denotes XOR operation, $\epsilon_0,\epsilon_1,\epsilon_2,a_i,b_{ij},
c_{ijk}$ are real number coefficients.
In the experiments, we set $M=8$ and each number $a_i$, $b_{ij}$ or $c_{ijk}$
is picked up from an uniform random variable  whose range is from $-1$ to $1$.
We regard last three components as the target variable $Y=(s_6,s_7,s_8)$
and the others as the feature variables $X_i=s_i,(i=1,..,5)$.

\begin{figure}[h!]
\centering
\includegraphics[width=0.9\hsize]{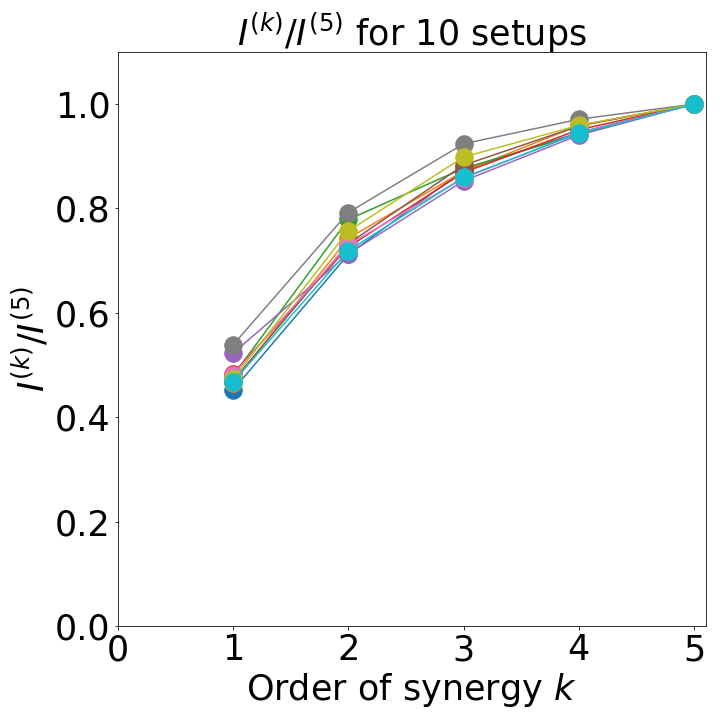}
\caption{
$I^{(k)}$ normalized by the total mutual information $I^{(5)}$
for 10 different setups.
}
\label{fig:Ik}
\end{figure}

In the first experiment, we calculate exact $I^{(k)}$.
We set $\epsilon_0=1,\epsilon_1=1/2,\epsilon_2=1/10$
in order to suppress higher order interactions.
By resampling coefficients $a_i,b_{ij},c_{ijk}$ from an uniform random variable from $-1$ to $1$,
we calculate $I^{(k)}$ for each setup.
Fig.~\ref{fig:Ik} shows results of $I^{(k)}$ normalized by $I^{(5)}$. Note that $I^{(5)}$ is equal to the total mutual information
and typically take values around $0.1-0.2$.
We take 10 different coefficient sets and plot them.
We can see that $I^{(k)}$ is actually increasing function.
In addition, upward convex curves of each result indicate the suppression of
information in higher order synergistic part.

\begin{figure}[h!]
\centering
\includegraphics[width=0.9\hsize]{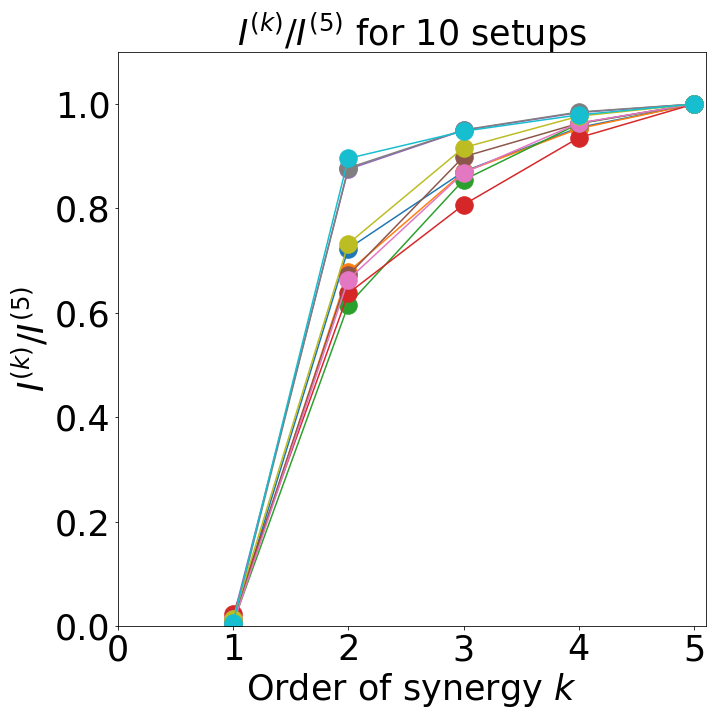}
\caption{
The strong coupling case of
$I^{(k)}$ normalized by the total mutual information $I^{(5)}$
for 10 different setups.
}
\label{fig:Ik_st}
\end{figure}

We also check the behavior of $I^{(k)}$ under the situation
where higher order interactions dominate over lower order ones.
In such a case, our approximation scheme is not expected to
work well. 
We set $\epsilon_0=1/10$, $\epsilon_1=1/100$ 
and $\epsilon_2=2$.
We take
 $a_i,b_{ij},c_{ijk}$ from an uniform random variable from $-1$ to $1$.
Then, we set $b_{ij}$ and $c_{ijk}$ that are not involved by just
one target variable to zero. 
We calculate $I^{(k)}$ for 10 setups.
Fig.~\ref{fig:Ik_st} shows results of $I^{(k)}$ normalized
by $I^{(5)}$. Since the dominant interaction terms are ones with $\epsilon_2$
involved by one target variable,
$I^{(2)}$ becomes much larger than $I^{(1)}$.
In this case, the leading order approximation $I^{(1)}$ does not work well.

\begin{figure}[h!]
\centering
\includegraphics[width=1.3\hsize]{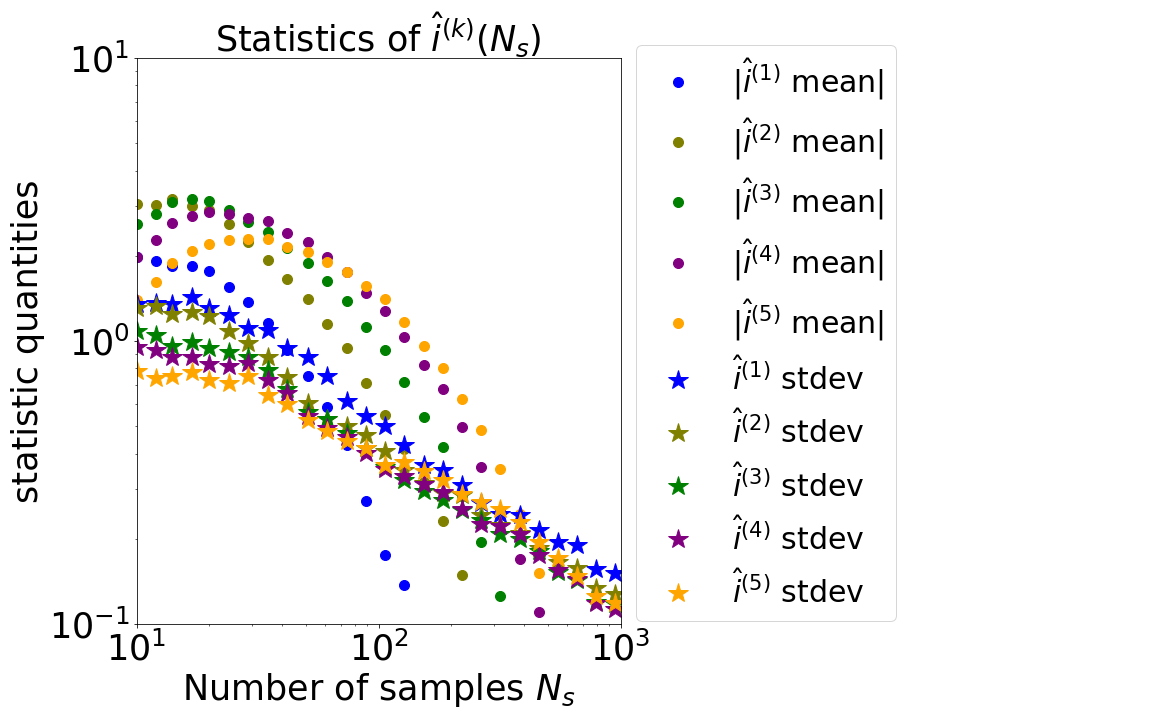}
\caption{
The mean and standard deviation of
$\hat{i}^{(k)}$ as a function of sample size $N_s$.
The term ``stdev'' is a shorthand notation of standard deviation.
}
\label{fig:stat}
\end{figure}

The second experiment is intended to see effects of finite sample size.
we set $\epsilon_0=1,\epsilon_1=1/2,\epsilon_2=1/10$ again.
We first fix the coefficients $a_i,b_{ij},c_{ijk}$ and calculate exact $I^{(k)}$.
Then, we pick up $N_s$ samples by using $p(s_1,..,p_M)$
and calculate the empirical probability $\hat{p}$.
In order to obtain $I^{(k)}$, we have to estimate $I(Y=y:c_i^{(k)})$ in Eq.~(\ref{eq:ik}).
We denote $\hat{I}(Y=y:C_i^{(k)})$
as one estimated by the empirical probability.
This $\hat{I}(Y=y:C_i^{(k)})$
is supposed to have relatively large bias especially
for small sample cases
\footnote{
It is known entropy related functions
have relatively large bias especially for small
sample cases.
In the literature,
a lot of sophisticated methods to
estimate bias term in entropy and mutual information
have been derived (see for
example~\cite{jisaku2}).
Since
derivation of precise bias correction is beyond the scope of this paper,
we use a simple bias correction in Eq.~(\ref{eq:bias}).}.
At the leading order in
asymptotic expansion, the bias 
 $\delta(y,C_i^{(k)})$ 
 would be given by
\begin{align}
\label{eq:bias}
\hat{p}(y)\delta(y,C_i^{(k)})
&=\sum_{c_v}\frac{1}{2N_s}
[1-\hat{p}(y,c_v)]\nonumber \\
&+\sum_{c_v}\frac{\hat{p}(y,c_v)}{2N_s\hat{p}(y)}
[1-\hat{p}(y)]\nonumber \\
&+\sum_{c_v}\frac{\hat{p}(y,c_v)}{2N_s\hat{p}(c_v)}
[1-\hat{p}(c_v)],
\end{align}
where $c_v$ denotes one of the possible values of $C_i^{(k)}$.
In App.~\ref{app:bi}, we derive this bias correction term.
We use bias corrected quantities
\begin{align}
\hat{I}(Y=y,C_i^{(k)})\rightarrow
\hat{I}(Y=y,C_i^{(k)})-\delta(y,C_i^{(k)})
\end{align}
in the estimation of $I^{(k)}$
and the result is denoted by $\hat{I}^{(k)}$.
We define a normalized variable as follows.
\begin{align}
\hat{i}^{(k)}=\frac{\hat{I}^{(k)}}{I^{(k)}}-1.
\end{align}
$\hat{i}^{(k)}$ can be regarded as a random variable against 
resamplings.
For a fixed sample size $N_s$,
we estimate the mean and standard deviation of $\hat{i}^{(k)}$ that
are functions of $N_s$.
In the estimation of mean and standard deviation, we take 100 time resamplings.
Fig.~\ref{fig:stat} shows mean and standard deviation of
$\hat{i}^{(k)}$ as a function of sample size $N_s$.
The mean value can be regarded as a bias from the 
true value.
In this setup with relatively small sample size,
we observe that the bias dominates over standard deviation.
We can see that lower order $I^{(k)}$ has less bias and stable when sample size is relatively small.

\section{Discussion}
\label{sec:dis}
In this paper, we have derived an approximation scheme
for multivariate information based on partial information
decomposition.
The key assumption is that information only appearing
in higher order synergy is small and
we have constructed truncation scheme for multivariate
information in terms of synergistic order.
The resulting approximation scheme is expected to be
reasonable when the higher order information in the system  
is suppressed. In addition, it is calculable in practice
when the truncation order is not so high.
We have also checked properties of our approximation scheme
by numerical experiments.

The truncated mutual information $I^{(k)}$ has
relatively simple structure. This simplicity
is originated from the simple structure of
the redundant information $I_{\rm min} (A_1,..,A_k)$.
Though the quantity $I_{\rm min}$ itself is well defined,
$I_{\rm min}$ does not take joint properties between
$A_i$ into account and it would overestimate the redundant
information in some sense. 
Due to this overestimation,
$I^{(k)}$ might underestimate the corresponding information
and lead to unintuitive results in some cases.
Thus,
it might be interesting to define an approximation scheme
based on another kind of redundant information $I_\cap$.

One direction of application of our approximation scheme
would be feature selection in machine learning.
Given a truncation order, we can
see important features in the system based on $I^{(k)}$.
Some of features may not contribute to $I^{(k)}$
and we obtain a minimal set of features that contribute
to $I^{(k)}$.
As is mentioned above, $I^{(k)}$ potentially underestimate the
information, which could cause underestimation of
the number of relevant features.
In this point of view, the feature selection based on $I^{(k)}$
can be regarded as a conservative one.
In any case, it would be interesting to see validity of feature selection
based on $I^{(k)}$ in realistic setups and future research
will be focused on it.
\section*{Acknowledgements}
We thank Daigo Honda and Nobuhiro Yonezawa for helpful
discussions and comments. We also thank KKST team
in Linea Co.,Ltd. for motivating and encouraging this study.

%
%
\bibliographystyle{unsrt}
\bibliography{ref}


\appendix
\section{Derivation of bias term}
\label{app:bi}
Here, we derive a bias correction term in Eq.~(\ref{eq:bias}).
We consider a bias correction for the quantity
$\hat{p}(y)\hat{I}(Y=y,C_i^{(k)})$.
This quantity can be rewritten as follows.
\begin{align}
\label{eq:aap}
\hat{p}(y)\hat{I}(Y=y,C_i^{(k)})=
\sum_{c_v \in C_i^{(k)}}\hat{p}(y,c_v)
\log\left(
\frac{\hat{p}(y,c_v)}{\hat{p}(y)\hat{p}(c_v)}
\right),
\end{align}
where $\hat{p}(\cdot)$ denotes empirical
probability.
For each probability $\hat{p}(\cdot)$, we define deviation term
$\Delta (\cdot)$ as follows.
\begin{align}
\Delta (\cdot)\equiv \frac{\hat{p}(\cdot)-p(\cdot)}{p(\cdot)},
\end{align} 
where $p(\cdot)$ denotes the true probability.
We expand Eq.~(\ref{eq:aap}) up to second order
in terms of $\Delta$
and calculate the average of it.
In the average calculation, we use properties of
the multinomial distribution.
Then, the result is given by Eq.~(\ref{eq:bias}).

\end{document}